\documentstyle[12pt]{article}

\catcode `\@=11
\@addtoreset{equation}{section}
\def\theequation{\arabic{section}.\arabic{equation}}
\catcode `\@=12

\newcommand{\be}{\begin{equation}}
\newcommand{\en}{\end{equation}}
\newcommand{\bea}{\begin{eqnarray}}
\newcommand{\ena}{\end{eqnarray}}
\newcommand{\beano}{\begin{eqnarray*}}
\newcommand{\enano}{\end{eqnarray*}}
\newcommand{\bee}{\begin{enumerate}}
\newcommand{\ene}{\end{enumerate}}

\newcommand{\R}{R \!\!\!\! R}
\newcommand{\N}{N \!\!\!\!\! N}

\newcommand{\Hil}{{\cal H}}
\newcommand{\I}{{\cal I}}

\newcommand{\Id}{1\!\!1}
\newcommand{\F}{{\cal F}}
\newcommand{\Lc}{{\cal L}}
\newcommand{\M}{{\cal M}}
\newcommand{\C}{{\cal C}}
\newcommand{\B}{{\cal B}}
\newcommand{\A}{{\cal A}}
\newcommand{\G}{{\cal G}}
\newcommand{\D}{{\cal D}}

\begin{document}

\thispagestyle{empty}

\vspace*{1cm}

\begin{center}
{\Large \bf Fixed Points in Topological *-Algebras of Unbounded
Operators}

\vspace{2cm}

{\large Fabio Bagarello}
\footnote[1]{Dipartimento di Matematica ed Applicazioni,
Fac. Ingegneria, Universit\`a di Palermo, I-90128  Palermo, Italy. e-mail: bagarell@unipa.it
Communicated by T. Kawai, July 14, 2000. Revised January 16, 2001}
\vspace{2mm}\\
\end{center}

\vspace*{2cm}

\begin{abstract}
\noindent
We discuss some results concerning fixed point equations in the setting of
topological
*-algebras of unbounded operators. In particular, an existence result is obtained for what
we have  called {\em
weak $\tau$ strict contractions}, and some continuity properties of these
maps are discussed. We
also discuss possible applications of our procedure to quantum mechanical
systems. \end{abstract}

\vspace{5cm}

{\bf Running title}: Fixed Points  in Topological *-Algebras.

\vfill

\newpage

\section{Introduction and Mathematical Framework}

Fixed point theorems have often proved to be powerful tools for abstract
analysis as well
as for concrete applications, see \cite{smart,rs,kant} for general overviews. In particular,
contraction mappings have been
successfully used in quantum mechanics for the description of systems with
infinite degrees of freedom,
$QM_\infty$, see \cite{rs}, Section V.6.c, and \cite{bm}. In this last reference, for instance, the
existence of an
(unique) fixed point has been used in the analysis of the thermodynamical
limit of (a class of) mean
field spin models.

On a  different side, it is well known to all the people
working on the 
algebraic approach to $QM_\infty$, \cite{hk}, that $C^*$ or Von Neumann
algebras are not reach
enough to be useful in the description of many physically relevant systems.
 For instance, difficulties already arise  in ordinary quantum
mechanics, since the commutation rule $[x,p]=i$ implies that the
operators $x$ and $p$ cannot be both bounded as operators on $\Lc^2(\R)$.
 These physical
difficulties have originated a wide
literature about unbounded operators and possible extensions of
$C^*$-algebras:  quasi *-algebras, \cite{las},  partial
*-algebras, \cite{ant},
 $CQ^*$-algebras,
\cite{btcq}, etc.. .  As for physical applications of these structures to $QM_\infty$, some are given in
\cite{bag,bag2,bt,bt2}.

In view of these considerations, it is natural to extend the notion of
contraction mappings to
quasi *-algebras, and then consider the consequences of this extension.

\vspace{3mm}

Before giving our definition of contraction mappings and in order to keep the paper self-contained, we briefly review some relevant definitions concerning quasi *-algebras.

Let $\Hil$ be a Hilbert space and $N$ an unbounded, self adjoint operator
defined on a  dense
domain $D(N)\subset \Hil$. Let  $D(N^k)$ be the domain of the operator
$N^k$, $k\in \N$, and
$\D$  the domain of all the powers of $N$:
\be
\D \equiv D^\infty(N) = \cap_{k\geq
0} D(N^k). \label{22}
\en
To be concrete we take here $N$ as the number operator for bosons, $N=a^\dagger a$, $a$ and $a^\dagger$ being the annihilation and creation operators satisfying the commutation relation $[a,a^\dagger]=\I$.

$\D$ is dense in $\Hil$. Following
Lassner, \cite{las}, we define the *-algebra $\Lc^+(\D)$  of all
the closable operators defined on $\D$ which, together with their
adjoints, map $\D$ into itself. It is clear that all  the powers of $a$ and
$a^\dagger$ belong to this set.

We define on $\D$ a topology $t$ 
by means of the following  seminorms:
\be \phi \in \D \rightarrow \|\phi\|_n\equiv \|N^n\phi\|,
\label{23} \en
where $n$ is a natural integer and $\|\, \|$ is the norm of $\Hil$, \cite{las}. The
topology $\tau$ in $\Lc^+(\D)$ is given as follows: we start
introducing the set  $\C$ of all the positive, bounded and continuous
functions $f(x)$ on $\R_+$, such that
\be
\sup_{x\geq 0}f(x)x^k<\infty, \quad \quad \forall k \in \N.
\label{23bis}
\en
 The seminorms on  $\Lc^+(\D)$ are labeled by the functions of the set
$\C$ and by the natural numbers $\N$. 
Therefore
$\|\:\|^{f,k}$ is a
seminorm of the topology $\tau$ if and only if $(f,k)$ belongs to the set $\C_N:=\{(\C,\N)\}$. We have
\be X\in \Lc^+(\D)
\rightarrow \|X\|^{f,k} \equiv
\max\left\{\|f(N)XN^k\|,\|N^kXf(N)\|\right\}, \quad \quad (f,k)
\in\C_N.  \label{24}
\en
Here $\|\,\|$ is the usual norm in $B(\Hil)$. From this definition
it follows that:
$\|X\|^{f,k}=\|X^\dagger\|^{f,k}$. In \cite{las} it has also
been proved that  $\Lc^+(\D)[\tau]$ is a complete locally convex
topological *-algebra.

Let us remark now that the two
contributions in the
definition (\ref{24}) have exactly  the same form. It is clear that, therefore, the estimate of
$\|f(N)XN^k\|$ is
quite close to that of $\|N^kXf(N)\|$, for any given $X\in
\Lc^+(\D)$. This is why we will 
identify $\|X\|^{f,k}$ with
$\|f(N)XN^k\|$ in the
following.

Moreover, using the spectral decomposition for $N$, $N=\sum_{l=0}^\infty l
\Pi_l$, the seminorm $\|X\|^{f,k}$ can be written as follows,  \cite{las}:
 \be
X\in \Lc^+(\D) \longrightarrow \|X\|^{f,k} = \sum_{l,s=0}^{\infty}
f(l)s^k\|\Pi_lX\Pi_s\|.
\label{25}
\en

\vspace{4mm}

The paper is organized as follows:

\noindent
in the next Section we introduce the notion of weak $\tau$ strict contractions and discuss the existence (and the uniqueness) of a fixed point for these maps;
in Section III we discuss the case in which the generalized contractions depend continuously on a parameter;
Section IV is devoted to examples and applications to differential equations, to ordinary quantum mechanics and to QM$_\infty$. The outcome is contained in Section V.
In the Appendix we will introduce, for practical convenience, a different topology $\tau_0$, equivalent to
$\tau$ and prove the non triviality of our construction. Of course $\Lc^+(\D)[\tau_0]$ is again a complete locally convex topological *-algebra.

\section{The Weak $\tau$-Strict Contractions}

Let $\B$ be a $\tau$-complete subspace of $\Lc^+(\D)$ and $T$ a map from
$\B$ into $\B$. We
say that $T$  is a {\em weak $\tau$ strict contractions over $\B$}, briefly
w$\tau$sc$(\B)$, if  there exists a constant $c \in]0,1[$ such that, for all
$(h,k)\in\C_N$, it
exists a pair $(h',k')\in\C_N$ satisfying \be \|Tx-Ty\|^{h,k}\leq c
\|x-y\|^{h',k'}\hspace{3cm}\forall \: x,y \in \B.  \label{31}
\en

As in the standard situation, see \cite{smart,rs,kant}, this definition
does not imply that
$\|Tz\|^{h,k}\leq c \|z\|^{h',k'}$ for all $ z\in \B$ since $T$ is not a linear map in general. Of course,
because  of this lack of linearity,  $T0$ could be different
from $0$; however, any such $T$ defines in a natural way another map $T'$
which is still a
w$\tau$sc$(\B)$ corresponding to  the same quantities $c$, $h'$ and $k'$ as the original map $T$ and which satisfies
$T'0=0$. In fact, let us put
 $T'x:=Tx-T0$, for all $ x\in \B$. Obviously we have $T'0=0$, and
$\|T'x-T'y\|^{h,k}=\|Tx-Ty\|^{h,k}\leq c \|x-y\|^{h',k'}$ for all choices
of $ x,y \in \B$. If $T0=0$, equation (\ref{31}) implies
 that \be \|Tx\|^{h,k}\leq c
\|x\|^{h',k'}\hspace{3cm}\forall x \in \B. \label{31bis} \en

In what follows we will consider equations of the form $Tx=x$, $T$ being a w$\tau$sc($\B$). The
first step consists
in introducing the following subset  of $\B$: \be \B_L\equiv\left\{x\in \B:
\sup_{(h,k)\in\C_N}\|Tx-x\|^{h,k}\leq L\right\}, \label{313}
\en
$L$ being a fixed positive real number. It is clear that, if
$L'>L$, then
$\B_L\subset \B_{L'}$. Some of the properties of these sets are contained
in the following

\vspace{3mm}

{\underline {\bf Lemma 1}}:-- Let $T$ be a w$\tau$sc($\B$). Then

(a) if $T0=0$ then any $x\in \B$ such that $\sup_{(h,k)\in\C_N}\|
x\|^{h,k}\leq L_1$ belongs to
$\B_L$ for $L\geq L_1(1+c)$;

(b) if $\|T0\|^{h,k}\leq L_2$ for all $(h,k)\in\C_N$, then any $x\in \B$
such that
$\sup_{(h,k)\in\C_N}\| x\|^{h,k}\leq L_1$ belongs to
$\B_L$ for $L\geq L_1(1+c)+L_2$;

(c) if $x\in \B_L$ then $T^n x\in \B_L$, for all $n\in \N$;

(d) $\B_L$ is $\tau$-complete;

(e) if $\B_L$ is not empty, then $T$ is a w$\tau$sc($\B_L$).

\vspace{3mm}

{\bf Proof}

(a) Due to the hypothesis on $\| x\|^{h,k}$ and to equation (\ref{31bis})
we have
\beano
\|Tx-x\|^{h,k}\leq&&\!\!\!\!\!\! \|Tx\|^{h,k}+\|x\|^{h,k}\leq
c\|x\|^{h',k'}+\|x\|^{h,k}\leq \\
\leq &&\!\!\!\!\!\! c\sup_{(h',k')\in\C_N
}\|x\|^{h',k'}+\sup_{(h,k)\in\C_N}\|x\|^{h,k}\leq
L_1(1+c).  \enano

(b) The proof uses the inequality
$$
\|Tx-x\|^{h,k}\leq \|Tx-T0\|^{h,k}+\|T0-x\|^{h,k},$$
together with (\ref{31}) for $y=0$ and the bound on $\|T0\|^{h,k}$.

(c) We prove the statement by induction. For $n=1$ we have
$$
\|T(Tx)-Tx\|^{h,k}\leq c \|Tx-x\|^{h',k'}\leq c \sup_{(h',k')\in\C_N
}\|Tx-x\|^{h',k'} \leq
cL\leq L, $$
Taking the $\sup_{(h,k)\in\C_N }$ of this inequality we conclude that
$Tx\in \B_L$. The second
step of the induction  goes as follows:
\beano
\|T(T^{n+1}x)- &&\!\!\!\!\!
T^{n+1}x\|^{h,k}=\|T(T^{n+1}x)- T(T^{n}x)\|^{h,k}\leq \\ &&\!\!\!\!\leq c
\|T^{n+1}x-
T^{n}x\|^{h',k'}\leq c \sup_{(h',k')\in\C_N }\| T(T^{n}x)- T^{n}x
\|^{h',k'}\leq cL\leq L,
\enano
 which implies that $ T^{n+1}x$ belongs to $\B_L$ whenever $ T^{n}x$ does.

(d) We will consider here the case in which $\B_L$ is non empty.  
Since $\B_L$ is a subset of a $\tau$-complete set, it is enough to check
that $\B_L$ is $\tau$-closed.
Let us take a sequence $\{x_n\}\in \B_L$, $\tau$-converging to an element
$x$. We  have to prove that
$x\in \B_L$.

First of all, it is evident that $T$ is $\tau$-continuous: in fact, if
$\{z_n\}$ is
$\tau$-convergent to $z$, then $\{Tz_n\}$ $\tau$-converges to $Tz$.
Moreover, since $x_n$
belongs  to $\B_L$ for all $n$, we have $\sup_{(h,k)\in\C_N
}\|Tx_n-x_n\|^{h,k}\leq L$
independently of $n$. We can conclude, therefore, that $$ \|Tx- x\|^{h,k}\leq
\lim_{n\rightarrow\infty}\|Tx_n- x_n\|^{h,k}\leq L, $$ which concludes the proof.

(e) This statement follows from the facts that $T$ is a w$\tau$sc($\B$),
that $T$ maps $\B_L$
into itself, and from the $\tau$-completeness of $\B_L$.

\hfill$\Box$

\vspace{4mm}

A consequence of this Lemma is that, if $\B_L$ contains a single element, then $\B_L$ is rather
a rich set.  What the Lemma does not say, is whether or not $\B_L$ contains
at least one element.
Of course, due to its definition, the answer will depend on the
explicit form of the
map $T$ and from the family of seminorms which define the topology. The
non-triviality of the
definition (\ref{313}) is proved in the Appendix.

We give now our main fixed-point result for a w$\tau$sc.

\vspace{3mm}

{\underline {\bf Proposition 2}}:-- Let $T$ be a w$\tau$sc($\B$). Then

(a) $\forall x_0\in \B_L$ the sequence $\{x_n\equiv T^nx_0\}_{n\geq 0}$
 is $\tau$-Cauchy in $\B_L$. Its $\tau$-limit, $x\in \B_L$, is a fixed
point of $T$;

(b) if $x_0,y_0\in \B_L$ satisfy the condition
$\sup_{(h,k)\in\C_N}\|x_0-y_0\|^{h,k}<\infty,$
then $\tau-\lim_nT^nx_0=\tau-\lim_nT^ny_0$.

\vspace{3mm}

{\bf Proof}

(a) First we observe that, due to the definition of $\B_L$, we have
$$
\|x_{n+1}-x_n\|^{h,k}=\|Tx_{n}-Tx_{n-1}\|^{h,k}\leq
c\|x_{n}-x_{n-1}\|^{h_1,k_1}\leq......\leq c^n
\|Tx_{0}-x_{0}\|^{h_n,k_n}\leq Lc^n,
$$
which implies, for any $n>m$, 
$$
\|x_{n+1}-x_m\|^{h,k}\leq L\frac{c^m}{1-c},
$$
which goes to zero for $m$ (and $n$) diverging. Therefore the sequence $\{x_n\}_{n\geq 0}$ is $\tau$-Cauchy. Since $\B_L$ is $\tau$-complete,
see Lemma 1, there exists an element $x\in \B_L$ such that
$x=\tau-\lim_nT^nx_0$. Now, the
$\tau$-continuity of $T$ implies that $x$ is a fixed point. In fact:
$$
Tx=T\left(\tau-\lim_nT^nx_0\right)=\tau-\lim_nT^{n+1}x_0=\tau-\lim_nx_{n+1}=x.
$$

(b) let us call $x=\tau-\lim_nT^nx_0$ and $y=\tau-\lim_nT^ny_0$. Then,
using $n$ times inequality (\ref{31}), we get
$$
\|x-y\|^{h,k}=\lim_n\|T^nx_0-T^ny_0\|^{h,k}\leq \lim_n c^n\|x_0-y_0\|^{h_n,k_n}\leq \left[\sup_{(h,k)\in\C_N}\|x_0-y_0\|^{h,k}\right]\lim_n c^n=0,
$$
for all seminorms. Therefore $x=y$.
\hfill$\Box$

\vspace{3mm}

{\bf Remarks:}

1) The first remark is that if $x_0$ and $y_0$ are two operators of
$\Lc^+(\D)$ satisfying
the bounds $\sup_{(h,k)\in\C_N}\|x_0\|^{h,k}=L_{x_0}$ and
$\sup_{(h,k)\in\C_N}\|y_0\|^{h,k}=L_{y_0}$ then, if $T0=0$, both $x_0$ and
$y_0$ belong to $\B_L$ for
$L=\max(L_{x_0}(1+c), L_{y_0}(1+c))$ as a consequence of Lemma 1. Moreover, it
is easy to check that $x_0$ and $y_0$ satisfy the condition in (b) of the
Proposition, and, for
this reason, they produce the same fixed point. To this same conclusion we arrive even if $T0\neq 0$ but $\|T0\|^{h,k}\leq L_2$ for a positive
constant $L_2$, independent of $(h,k)$.

2) The second remark concerns the non uniqueness of the fixed point given by our
procedure. In fact, the  statement (b) above implies uniqueness only within
a certain class of
possible fixed points, those obtained starting from elements of $\B_L$. It is useless to stress that other possibilities could exist for
finding completely
different fixed points which are not considered here. However, there exists a simple situation in which
uniqueness is also ensured: it
happens  when the map $T$ is a {\em $\tau$ strict contraction over \B}. Such a map
differs from a
w$\tau$sc$(\B)$ in that the new seminorm $\|\|^{h',k'}$ in the r.h.s. of inequality
(\ref{31}) coincides with
the original one: \be \exists c\in ]0,1[ \mbox{ such that }
\|Tx-Ty\|^{h,k} \leq c \|x-y\|^{h,k}
\:
\forall x, y\in \B \mbox{ and } \forall (h,k)\in \C_N. \label{314} \en For
such a contraction everything
is much easier since a standard result, see \cite{rs}, Theorem V.18,
can be adapted here without
major changes, and gives the existence and uniqueness of the fixed point.

3) The fact that the fixed point of the map $T$ belongs to $\B_L$ should
not be a big
surprise. As a matter of fact, this is true just because of the definition of fixed point. In fact, since $Tx-x=0$, it is clear that
$\sup_{(h,k)\in\C_N}\| Tx-x \|^{h,k}=0$, which implies that $x\in \B_L$.

4) This fixed point result is  different  from the one given in \cite{rs}, Theorem V.18, where the sequence $\{T^nx\}$ could be constructed starting by any element $x$ of the complete metric space on which $T$ acts. Here, on the other way, the role of $\B_L$, as the set of the {\em starting points} for the sequences $\{x_n\}_{n\geq 0}$ producing the fixed points, is crucial!.

5) We also remark  that the hypothesis of Proposition 2, point (b), is
verified by the pair
$(x_0,y_0\equiv T^mx_0)$, $m$ being a fixed natural, so that the related
fixed points coincide. This
is, again, not surprising since, of course, the two sequences $\{
T^nx_0\}_{n\geq 0}$ and $\{
T^n(T^mx_0)\}_{n\geq 0}$ must converge to the same element.

6) The same
procedure can be  generalized to the  situation in which
$\Lc^+(\D)$ is
replaced by an algebra (or a *-algebra, if needed) $\A$, which is complete
with respect to a
locally convex topology $\sigma$ defined by a (non countable) family of
seminorms $p_\alpha$.
In this case, if $\B$ is a  $\sigma$-complete subspace of $\A$ and $T$ is a
map from $\B$ into
$\B$ we say that $T$ is a w$\sigma$sc($\B$) if there exists a constant
$c\in]0,1[$ such that, for any
seminorm $p_\alpha$, there exists a (different) seminorm $p_\beta$ for which
$$ p_\alpha(Tx-Ty)\leq
c p_\beta(x-y), \hspace{3cm}\forall x,y\in \B. $$
Lemma 1 and Proposition 2 can be stated with only minor changes.

\section{Continuity of Weak $\tau$-Strict Contractions}

In this Section we consider the case in which the w$\tau$sc($\B$) depends
on a (real) parameter
assuming that some kind of continuity  holds. Besides its mathematical
interest,  this situation
has a certain relevance in quantum mechanics, which will be discussed in the next Section.

Let $I\subset \R$ be a set such that $0$ is one of its accumulation points.
A family of weak $\tau$ strict contractions $\{T_\alpha\}_{\alpha\in I}$ is
said {\em uniform} if

1) $T_\alpha:\B \rightarrow \B \:\: \forall \alpha\in I$, $\B$ being a
$\tau$-complete subspace of $\Lc^+(\D)$;

2) $\forall (h,k)\in \C_N$ and $\forall \alpha \in I$ there exist $(h',k')
\in \C_N$,  independent of
$\alpha$, and  $c_\alpha\in ]0,1[$, independent of $(h,k)$, such that \be
\|T_\alpha x- T_\alpha y\|^{h,k}\leq c_\alpha \|x- y\|^{h',k'}, \quad
\forall x,y \in \B;
\label{42}
\en

3) $c_-\equiv \lim_{\alpha, 0}c_\alpha \in]0,1[$.

Again, it is worthwhile to remark that none of the $T_\alpha$ is supposed
to be linear, so
that $ T_\alpha x- T_\alpha y$ needs not to coincide with $T_\alpha(x-y)$.

An important consequence of this uniformity is that the element $(h_n,k_n)\in \C_N$ in the r.h.s. of the inequality below is independent of the order of the maps $T_\alpha$  and only
depends on the initial pair $(h,k)$ and on the number of maps, $n$:   
\be
\|T_{\alpha_1} T_{\alpha_2} .....T_{\alpha_n} x
\|^{h,k}\leq c_{\alpha_1} c_{\alpha_2} .....c_{\alpha_n} \|x \|^{h_n,k_n}.
\label{43}
\en

We further say that the family $\{T_\alpha\}_{\alpha\in I}$ is
$\tau$-strong Cauchy if, for
all $(h,k)\in \C_N$ and $\forall y\in\B$,
 \be
\|T_\alpha y- T_\beta y\|^{h,k}\rightarrow 0, \quad
\label{44}
\en
whenever both $\alpha$ and $\beta$  go to zero.

With natural notation, we call $\B_L^{(\alpha)}$ the set $\B_L$ related to
the map $T_\alpha$,
$$\B_L^{(\alpha)}\equiv\left\{x\in \B:
\sup_{(h,k)\in\C_N}\|T_\alpha x-x\|^{h,k}\leq L\right\}.$$
We stress that, even if the set $\B$ is unique for
all the maps $T_\alpha$, 
the sets
$\B_L^{(\alpha)}$ may differ from each other.

\vspace{3mm}

{\underline {\bf Proposition 3}}:-- Let $\{T_\alpha\}_{\alpha\in I}$ be a
$\tau$-strong
Cauchy uniform family of w$\tau$sc($\B$). Then

1) There exists a w$\tau$sc($\B$), $T$, which satisfies the following relations:
\be
\|T y- T_\alpha y\|^{h,k}\rightarrow 0 \quad \forall y\in\B, \, \forall
(h,k)\in \C_N
\label{45}
\en
and
\be
\|T y- T z\|^{h,k}\leq c_-\| y- z\|^{h',k'}\quad \forall y,z\in\B,
\label{46}
\en
where $( h',k')$ are those of inequality (\ref{42}).

2) let $\{x_\alpha\}_{\alpha\in I}$ be a family of fixed points of the net
$\{T_\alpha\}_{\alpha\in I}$: $ T_\alpha x_\alpha=x_\alpha$, $\forall
\alpha\in I$. If $\{x_\alpha\}_{\alpha\in I}$
is a $\tau$-Cauchy net then, calling $x$ its $\tau$-limit in $\B$,  $x$ is
a fixed point of
$T$.

3) If the set $\cap_{\alpha\in I} \B_L^{(\alpha)}$ is not empty and if the
following
commutation rule holds

\be
T_\alpha (T_\beta y) = T_\beta (T_\alpha y), \quad \forall \alpha,\beta\in
I  \mbox{ and }
\forall y\in \B, \label{47}
\en
 then, calling
\be
x_\alpha=\tau-\lim_{n\rightarrow \infty} T_\alpha^nx^0 \quad x^0\in
\cap_{\alpha\in I} \B_L^{(\alpha)},
\label{48}
\en
each $x_\alpha$ is a fixed point of  $T_\alpha$,  $T_\alpha
x_\alpha=x_\alpha $ and $\{x_\alpha\}_{\alpha\in I}$ is a $\tau$-Cauchy
net. Moreover
$\tau-\lim_{\alpha\rightarrow 0} x_\alpha$ is a fixed point of $T$.

\vspace{3mm}

{\bf Proof}

1) Since $\B$ is $\tau$-complete and since $\{T_\alpha\}_{\alpha\in I}$ is
$\tau$-strong Cauchy, for any $y\in\B$ there exists an element $z\in \B$ such
that $z=\tau-\lim_{\alpha, 0} T_\alpha y$. We use $z$ to define $T$ as
\be
Ty:=z.
\label{49}
\en
It is evident that $T$ maps $\B$ into itself and that $\|T y- T_\alpha
y\|^{h,k}\rightarrow 0$ for every $y\in\B$ and for all $(h,k)\in \C_N$.

Equation (\ref{46}) follows from:
$$
\|T y- Tz\|^{h,k}=\lim_{\alpha\rightarrow 0}\|T_\alpha y- T_\alpha
z\|^{h,k}\leq
\lim_{\alpha\rightarrow 0}c_\alpha \| y-  z\|^{h',k'}=c_-\| y-  z\|^{h',k'},
$$
for all $y,z\in \B$. 

\vspace{3mm}

2) Since $\{x_\alpha\}_{\alpha\in I}$ is $\tau$-Cauchy, there exists in $\B$
an element $x=\tau-\lim_\alpha
x_\alpha$. We use the equality $ T_\alpha x_\alpha=x_\alpha $ to  prove
that $ Tx=x$.
In fact $$ \|T x- x\|^{h,k}\leq
\|T x- T_\alpha
x\|^{h,k}+\|T_\alpha x-x_\alpha\|^{h,k}+\|x_\alpha- x\|^{h,k}\rightarrow 0. $$
This is because all the contributions in the rhs goes to zero for $\alpha$ going to
zero:  the first
because of the equation (\ref{45}), the third because of the definition of
$x$ and the second 
for the same reason, since $$\|T_\alpha x-x_\alpha \|^{h,k}=\|T_\alpha x-
T_\alpha
x_\alpha\|^{h,k}\leq c_\alpha \|x-x_\alpha \|^{h,k}.$$

\vspace{3mm}

3) Since it exists an element $ x^0\in \cap_{\alpha\in I} \B_L^{(\alpha)}$,
Proposition 2 implies
that $x_\alpha=\tau-\lim_{n \rightarrow \infty} T_\alpha^nx^0$ is a fixed
point of $ T_\alpha$,
and $x_\beta=\tau-\lim_{ n \rightarrow \infty } T_\beta^nx^0$ is a fixed
point of $ T_\beta$.
This means, using the definition of the limit, that for any fixed
$\epsilon>0$ and for each
$(h,k)\in\C_N$, there exists an integer $m$ such that \be \| T_\alpha^m
x^0-x_\alpha \|^{h,k}+\|
T_\beta^m x^0-x_\beta \|^{h,k}\leq \frac{2\epsilon}{3}. \label{410} \en
For this fixed $m$ we now estimate $\| T_\alpha^m x^0- T_\beta^m x^0
\|^{h,k}$. It is possible to
show that the following inequality holds: \be
\| T_\alpha^m x^0- T_\beta^m x^0 \|^{h,k}\leq m \| T_\alpha x^0- T_\beta
x^0 \|^{h_{m-1},k_{m-1} }
\label{411}
\en
where $( h_{m-1},k_{m-1})$ only depends on the origin pair $( h,k)$ and on
$m$, but not on
$\alpha$ and $\beta$. We prove this inequality only for $m=2$. Its
generalization to larger values of $m$ is straightforward. \beano
&&\!\!\!\!\!\| T_\alpha^2 x^0- T_\beta^2 x^0 \|^{h,k} \leq \|
T_\alpha(T_\alpha x^0)- T_\alpha
(T_\beta x^0) \|^{h,k}+\| T_\alpha (T_\beta x^0)- T_\beta (T_\beta x^0)
\|^{h,k}=\\ &&\!\!\!=\|
T_\alpha(T_\alpha x^0)- T_\alpha (T_\beta x^0) \|^{h,k}+\| T_\beta
(T_\alpha x^0)- T_\beta
(T_\beta x^0) \|^{h,k}\leq \\
&&\!\!\!\!\leq (c_\alpha+c_\beta) \| T_\alpha x^0- T_\beta x^0 \|^{h_1,k_1}\leq
 2\| T_\alpha x^0- T_\beta x^0 \|^{h_1,k_1}. \enano Here we have used condition
(\ref{47}) together with the remark leading to inequality (\ref{43}).

Now we can collect all these results to prove the statement: let $m$ be the
fixed
integer introduced in equation (\ref{410}). We have \beano
\| x_\alpha- x_\beta \|^{h,k} \leq  &&\!\!\!\!\!\| x_\alpha- T_\alpha^mx^0
\|^{h,k}+\|
T_\alpha^mx^0- T_\beta^mx^0 \|^{h,k}+\| T_\beta^mx^0- x_\beta \|^{h,k}\leq\\
&&\!\!\!\leq  \frac{2\epsilon}{3}+m\| T_\alpha
x^0- T_\beta x^0 \|^{h_{m-1},k_{m-1}}. \enano

Since $\{T_\alpha\}_{\alpha\in I}$ is $\tau$-strong
Cauchy then there exists a ball, $P(0,\gamma)$, centered in zero and with
radius $\gamma$, which depends
on $(h_{m-1},k_{m-1})$, $\epsilon$ and $m$, such that, for all $\alpha$ and
$\beta$ inside this
ball, the inequality $\| T_\alpha x^0- T_\beta x^0 \|^{h_{m-1},k_{m-1}}\leq
\frac{\epsilon}{3m}$ holds. In
conclusion we have proved that  $$
\forall \epsilon>0, \: \forall (h,k)\in\C_N \:\: \exists
P(0,\gamma) \mbox{ such that } \| x_\alpha- x_\beta \|^{h,k} \leq \epsilon,
\quad \forall
\alpha,\beta \in P(0,\gamma).  $$
This implies that $\{x_\alpha\}_{\alpha\in I}$ is a $\tau$-Cauchy net. The last
statement finally follows
from point 2).

\hfill$\Box$

\vspace{3mm}

We now consider a different kind of problem: let $\{T_\alpha\}_{\alpha\in
I}$ be an uniform family of w$\tau$sc($\B$), $\tau$-strong convergent to a
w$\tau$sc($\B$), $T$; let $x$ be a fixed point of $T$ and $x_\alpha$ a
fixed point of $T_\alpha$, $\alpha\in I$. We
wonder if the net $\{x_\alpha\}_{\alpha\in I}$ is $\tau$-converging to $x$.
Of course, sic
stantibus rebus, the answer cannot be positive, because of the non
uniqueness of the fixed
points of a generic w$\tau$sc.  In order to say something more we must impose  other conditions. We prove the following

\vspace{3mm}

{\underline {\bf Proposition 4}}:-- Let $\{T_\alpha\}_{\alpha\in I}$ be an
uniform family  of
w$\tau$sc($\B$), $\tau$-strong converging to the w$\tau$sc($\B$) $T$ and
satisfying condition
(\ref{47}).

If the set $(\cap_{\alpha\in I} \B_L^{(\alpha)}) \cap \B$ contains an
element $x^0$
then, defining 
$$
x_\alpha=\tau-\lim_{n\rightarrow \infty} T_\alpha^nx^0 \quad \mbox{ and } x
=\tau-\lim_{n\rightarrow \infty} T ^nx^0, \quad \quad \alpha \in I,$$
$\| x_\alpha- x \|^{h,k} \rightarrow 0$ for all $(h,k)\in\C_N$.

\vspace{3mm}

{\bf Proof}

First of all we observe that a consequence of condition (\ref{47}) is the
analogous
commutation rule for the maps $ T_\alpha$ and $T$: $$
T_\alpha (T y) = T (T_\alpha y), \quad \forall \alpha \in I \mbox{ and }
\forall y\in \B.
$$
Using this result, the statement follows from the same argument as the one used in the proof of Proposition 3,
point 3). \hfill$\Box$

\vspace{3mm}

An easier result can be obtained under stronger assumptions. First we
call
$\{T_\alpha\}_{\alpha\in I}$ an uniform family of $\tau$sc($\B$) if it is a
w$\tau$sc($\B$) and
if $(h',k')=(h,k)$, in inequality (\ref{42}). For any fixed $\alpha$ we have already observed in Section II that the fixed point of $T_\alpha$
is unique. Here we have:

\vspace{3mm}

{\underline {\bf Proposition 5}}:-- Let $\{T_\alpha\}_{\alpha\in I}$ be an
uniform family of $\tau$sc($\B$), $\tau$-strong converging to the
$\tau$sc($\B$) $T$, and such that $c_\alpha\leq c^+ \:\forall \alpha\in I$
with $c^+<1$.

If $\{x_\alpha\}_{\alpha\in I}\subset \B$ and $x\in \B$ are such that
$T_\alpha x_\alpha= x_\alpha$ for all $\alpha\in I$ and $Tx=x$, then
 $\| x_\alpha- x \|^{h,k} \rightarrow 0$ for all $(h,k)\in\C_N$.

{\bf Proof}

Under these hypotheses we have
\beano
&&\| x_\alpha- x \|^{h,k}= \| T_\alpha x_\alpha- Tx \|^{h,k}\leq \|
T_\alpha x_\alpha -
T_\alpha x \|^{h,k}+\| T_\alpha x- Tx \|^{h,k}\leq\\
 &&\leq c_\alpha \| x_\alpha- x \|^{h,k} +\| T_\alpha x- Tx \|^{h,k}\leq
c^+ \| x_\alpha- x
\|^{h,k} +\| T_\alpha x- Tx \|^{h,k}. \enano
Using the hypothesis on $c^+$ we conclude that $\| x_\alpha- x \|^{h,k}\leq
\frac{1}{1-c^+}\| T_\alpha x- Tx \|^{h,k}$, which goes to zero by
assumption.
\hfill$\Box$

\section{Examples and Applications}

We start this Section giving few examples of w$\tau$sc (we omit the space $\B$ whenever this
coincides with the
whole $\Lc^+(\D)$).

Using the same notations as in the Introduction, we define the following maps acting on  $x\in \Lc^+(\D)$:
$$
T_\alpha^{(i,j)}x\equiv \alpha N^ixN^j, 
$$
and
$$
T_lx\equiv [N^l,x]=N^lx-xN^l.
$$

Here $\alpha$ is a complex number with modulus strictly less than 1, while
$i,j$ and $l$ are natural numbers. We also assume that $N^{-1}$ exists (as a bounded operator) and satisfies the
 bound $2\| N^{-1}\|^l< 1$. Both $T_\alpha^{(i,j)}$ and $T_l$ are linear and it can be easily checked that they are w$\tau$sc. In fact, introducing the function $h_i(x)=x^ih(x)$, which still belongs to the set $\C$,  we get $\|T_\alpha^{(i,j)}x\|^{h,k}=|\alpha|\|x\|^{h_i,k+j}$ and $\|T_lx\|^{h,k}\leq 2\|
N^{-1}\|^l
\|x\|^{h_l,k+l}$. Our claim finally follows from the linearity of the maps.

\vspace{3mm}

More relevant is the following application to differential equations which shows that it is possible to associate a w$\tau$sc
to some differential
equations over $\Lc^+(\D)$.

Let $\delta$ be a positive real number, $\{d_{h,k}\}_{(h,k)\in\C_N}$ a net
of positive  real
numbers and $x_0$ an element of the algebra $\Lc^+(\D)$ (corresponding to the initial
condition).
Let us now introduce the following sets:
\be
I_\delta\equiv[0,\delta],
\label{33}
\en
\be
L_{x_0,\{d\},\delta}\equiv\{X\in \Lc^+(\D):\:\:\forall (h,k)\in\C_N \quad
\exists
(h',k')\in\C_N : \|X-x_0\|^{h,k}\leq \delta d_{h',k'} \}, \label{34} \en
and, finally,

\be
\F\equiv I_\delta \times L_{x_0,\{d\},\delta}.
\label{35}
\en
It is clear that this set is not empty; in fact, among other elements, it
contains $x_0$ for
any values of $\delta$ and for any choice of the net
$\{d\}\equiv\{d_{h,k}\}_{(h,k)\in\C_N}.$ (In order to simplify the
notation, we do not write the explicit dependence of $\F$ on $x_0,\{d\}$
and $\delta$.)

We further introduce the following set of functions: 
\bea
\M\equiv\{z(t):&&\!\!\!\!\!\! I_\delta \rightarrow \Lc^+(\D), \tau-\mbox
{continuous and such
that } \forall (h,k)\in \C_N \nonumber \\
&&\!\!\!\!\!\exists (h',k')\in \C_N : \|z(t)-x_0\|^{h,k}\leq \delta
d_{h',k'}\}.
\label{35bis}\ena

Let now $f(t,x)$ be a
function defined on $\F$ which takes values in $\Lc^+(\D)$, and  for which
a constant  $M$
exists, with   $0<M<\frac{1}{\delta}$,  such that for all $(h,k)\in C_N,$
there exist two pairs  $(h',k'), (h'',k'')\in C_N$ satisfying

\be
\| f(t,x)\|^{h,k}\leq  d_{h',k'}\hspace{4.2cm} \forall (t,x)\in \F
\label{37}
\en
\be
\| f(t,x)- f(t,y)\|^{h,k}\leq M \| x- y\|^{h'',k''}\hspace{3cm} \forall
(t,x),(t,y)\in \F.
\label{38}
\en
For such a function we consider the following differential equation
\be
\frac{dx(t)}{dt}=f(t,x(t)), \hspace{3cm} x(0)=x_0,
\label{39}
\en
which can be written in integral form as
\be
x(t)=x_0+\int_0^tds f(s,x(s)).
\label{310}
\en
Let us now introduce the following map $U$ on $\M$:
\be
(Uz)(t)\equiv x_0+\int_0^tds f(s,z(s)),
\label{311}
\en
$t\in I_\delta$.
It is obvious that, for a generic function $f(t,x)$, the map $U$ is not
linear and $U0\neq 0$.
 It can
be proven that $U$ is a w$\tau_\delta$sc($\M $), $\M$ being endowed with
the topology $\tau_\delta$
defined by the following seminorms: \be \|z \|^{h,k}_\delta:=\sup_{t\in
I_\delta}\|z(t)
\|^{h,k}. \label{312}
\en
The proof follows these lines:

first of all, it is easy to check that $\M$ is $\tau_\delta$-closed;

secondly, due to the bound (\ref{37}), we can verify that the function
$(Uz)(t)$ is
$\tau$-continuous. In fact, since  $z(s)$ belongs to $\M$, then $(s,z(s))$
belongs to $\F$. This
implies that, for all $(h,k)\in \C_N$ there exists another pair $(h',k')\in
\C_N$ such that $$
\|(Uz)(t)- (Uz)(t_0) \|^{h,k}\leq \int_{t_0}^t \| f(s,z(s)) \|^{h,k}ds\leq
d_{h',k'}|t-t_0|\rightarrow 0, $$ when $t\rightarrow t_0$. 

Yet, with analogous
estimates, we also
conclude  that $\|(Uz)(t)-x_0\|^{h,k}\leq \delta d_{h',k'}$ for all $t\in
I_\delta$, and that,
therefore, $U$ maps $\M$ into itself;

finally, condition (\ref{38})produces the
following estimate
$$
\|Uy-Uz \|^{h,k}_\delta \leq M\delta \|y-z \|^{h',k'}_\delta,
$$
for all $y,z$ in $\M$. Therefore, since $M\delta <1$,  $U$
is  a
w$\tau_\delta$sc($\M $).

\vspace{2.5mm}

Simple examples of functions satisfying conditions (\ref{37}) and
(\ref{38}) are:

(a) $f_1(t,x)=\varphi(t)\Id$, with $|\varphi(t)|\leq 1$ for all $t\in
I_\delta$. For this
example we fix the net $\{d\}$ as follows: $d_{h,k}:=\|\Id\|^{h,k}$, while
$x_0$ and $\delta$
are completely free;

(b) $f_2(t,x)=\varphi(t)X$, with $|\varphi(t)|\leq \frac{1}{2\delta}$ for
all $t\in I_\delta$
and $X\in \Lc^+(\D)$. Here we take for convenience $x_0=0$ while $\delta$
and $\{d\}$ are free;

(c) $f_3(t,x)=\varphi(t)N^lX$, with $|\varphi(t)|\leq \frac{1}{2\delta}$
for all $t\in
I_\delta$, $l\in\N$, $X\in \Lc^+(\D)$ and $N$ is the number operator introduced in Section I. Again, we fix $x_0=0$, while
$\delta$ and $\{d\}$ are free.

\vspace{3mm}

In order to apply Proposition 2 to the analysis of the differential equation (\ref{39})
we first have to check that the set $\B_L$ is non-empty. In other
words, it is necessary
to check that there exists (at least) an element $z_0(t)\in \M$ such that \be
\sup_{(h,k)\in\C_N}\|(Uz_0)(t)-z_0(t)\|^{h,k}_\delta\leq L,
\label{314bis}
\en
for a fixed positive constant $L$.

In general this check is not easy. However, if the function $f(t,x)$ and
the initial
condition $x_0$ satisfy, for a given $L'>0$, the estimate \be
\|f(t,x_0)\|^{h,k}\leq L', \hspace{3cm} \forall t\in I_\delta, \: \forall
(h,k)\in\C_N,
\label{315}
\en
then we can conclude that condition (\ref{314bis}) is verified by choosing
$z_0(t)\equiv x_0$ for all $t\in I_\delta$. In
fact, with this choice, 
we have $$
\|(Uz_0)(t)-z_0(t)\|^{h,k}_\delta=\sup_{t\in I_\delta}
\left\|\int_0^tf(s,x_0)ds\right\|^{h,k}\leq \sup_{t\in
I_\delta}\int_0^t\|f(s,x_0) \|^{h,k}
ds\leq L'\delta, $$ which implies the bound (\ref{314bis}).  In other words, if condition (\ref{315}) holds, $x_0$ can be
considered as a good
starting point to construct the solution of the differential equation.

A simple example in which condition (\ref{315}) is satisfied is $f(t,x)=\varphi(t)x$. Here $\varphi(t)$ is a "regular" function and
the initial condition
$x_0$ is defined using the  strategy used in the
Appendix in the proof of the non triviality of the set $\B_L$.

In conclusion we have shown that, under some conditions on the function $f(t,x)$ (which are not very different from the ones usually required in connection with the Cauchy problem), the existence, but not the uniqueness, of the solution of the differential equation (\ref{39}) follows from our results.

\vspace{3mm}

This procedure can be applied straightforwardly  to a generic quantum
mechanical dynamical problem.

Let $x$ be an element of the algebra $\Lc^+(\D)$ related to some quantum
mechanical  problem
whose time evolution we are interested in. For instance, we can think of
$\D$ as the domain of
all the powers of the number operator $N\equiv a^\dagger a$, $a$ and
$a^\dagger$ being
as in the Introduction. Let $H=N$ be the hamiltonian of
the system, which will
be used to construct the seminorms: $\| Y\|^{h,k}=\|h(H)YH^k\|$. The time
evolution $x(t)$ is
driven by the following Heisenberg equation: \be
\frac{dx(t)}{dt}=i[H,x(t)], \label{312bis}
\en
 with initial
condition $x(t_0)=x_0$.

It is well known  that a formal solution of this
equation does exist, and that its form is $x(t)= e^{iHt}x_0e^{-iHt}$. Now we want to show
that the existence of the solution of the equation (\ref{312bis}) can also be
obtained  by using our  analysis of abstract differential equations. In
particular, we will
prove  that to the Heisenberg equation of motion for an observable $x$
can be associated a
w$\tau$sc($\M$). 

Calling $f(t,x(t))= i[H,x(t)]$ we can write the differential equation
(\ref{312bis}) in the integral form $$
x(t)=x_0+\int_0^tf(s,x(s))ds. $$ It may be worthwhile to notice that this
is an example in which
the function $f(t,z)$ is linear in $z$ and  does not depend explicitly on
$t$. We now check that the function $f(t,x(t))$ satisfies
conditions (\ref{37}) and (\ref{38}). First of all, we define the net
$\{d\}$ by
$d_{h,k}:=\|x_0\|^{h,k}$. Secondly, fixed a  $\delta>0$, we introduce
the  sets defined in (\ref{33})-(\ref{35bis}). In the
following we will assume
also the following bound on $H^{-1}$: $$
\|H^{-1}\|\leq \frac{1}{2(\delta+1)}.
$$
Needless to say, this is not a strong assumption since, in any case, a
constant can  be
added to $H$ without affecting the equation of motion. We have, for
$(t,z)\in \F$,
\beano &&\!\!\| f(t,z)\|^{h,k}=\|h(H) f(t,z) H^k\|=
\|h(H) [H,z] H^k\|\leq \\ &&\leq \|h(H) H z H^k\|+\|h(H)z H^{k+1}\|\leq
2\|H^{-1}\|\|z\|^{h_1,k+1}, \enano

\noindent
where  $h_1(x)=xh(x)$. Now we use the fact that
$z$ belongs to
$L_{x_0,\{d\},\delta}$. This implies that, for all $(h,k)\in \C_N$, it
exists another element in
$\C_N$, $(h',k')$, such that $\|z-x_0\|^{h,k}\leq \delta d_{h',k'}$. Therefore
$\|z\|^{h_1,k+1}$ can be estimated by $\delta d_{h_1',(k+1)'}+d_{h_1,k+1}$.
Moreover, defining
$(h',k')$ as $(h_1',(k+1)')$ if $d_{h_1',(k+1)'}\geq d_{h_1,k+1}$ and as
$(h_1,k+1)$
otherwise, we conclude that $\|z\|^{h_1,k+1}\leq (\delta+1)d_{h',k'}$.
Therefore, recalling the
bound on $\|H^{-1}\|$, we find that
$$
\| f(t,z)\|^{h,k}\leq d_{h',k'},
$$
whenever  $(t,z)\in \F$, which is exactly condition (\ref{37}).

Moreover, for any $(t,z)$ and $(t,y)$ in $\F$,
 \beano
&&\!\!\! \| f(t,x)- f(t,y)\|^{h,k}\leq  \|h(H) H (x-y)
H^k\|+\|h(H)(x-y) H^{k+1}\|\leq \\
&&\!\!\!\leq 2\|H^{-1}\||h_1(H)(x-y) H^{k+1}\|=2\|H^{-1}\|\|x-y\|^{h_1,k+1}\leq
\frac{1}{1+\delta}\|x-y\|^{h_1,k+1}.  \enano
We conclude that both conditions on the function $f(t,x)$ are satisfied
since
$M\equiv\frac{1}{1+\delta}<\frac{1}{\delta},$ so that
 the map $U$ related to $H$ as in (\ref{311}) is a w$\tau$sc($\M$).

For what concerns the starting element which produces the fixed point, the situation is
again very close to that of general differential equations: if our
initial condition $x_0$ satisfies the bound $\|x_0\|^{h,k}\leq m$ for all
$(h,k)\in \C_N$,
then we can check that the choice $ z_0(t)=x_0$ for all $t\in I_\delta$
produces an element of
the set $\M_L:=\{y(t)\in \M: \,
\sup_{(h,k)\in\C_N}\|Uy-y\|^{h,k}_\delta\leq L\}$, for
$L=\frac{m\delta}{1+\delta}$, which can be used to construct the sequence
$\{U^nx_0\}_{n\in\N}$
$\tau_\delta$-converging to the solution of the Heisenberg equation.

\vspace{4mm}

We end this Section with another physical application, consequence of the results discussed in
Section III. We want to stress that now the
philosophy is rather different from that of 
the previous application where an existence result for the Heisenberg equation of motion
was deduced.
Here, on the other hand, we want to find the time evolution for a $QM_\infty$ system  in the thermodynamical limit.

 Let us consider a physical system whose energy is given by a certain
unbounded self-adjoint operator $H$, densely defined and invertible.
We define $\D$ to
be the domain of all the powers of the operator $H$, and $\Lc^+(\D)[\tau]$
the topological
*-algebra given in the Introduction. The seminorms are the usual ones,
$\|X\|^{h,k}=\|h(H)XH^k\|$, $(h,k)\in\C_N$. As widely discussed in the literature, the rigorous approach
to the physical
problem implies, as a first step, the introduction of a cut-off $\alpha$
which makes the model
well defined, and the related hamiltonian $H_\alpha$ a
self-adjoint
bounded operator. In what follows we will assume that $\alpha$ takes value
in a given subset
$I$ of $\R$ and that the limit $\alpha\rightarrow 0$ corresponds to the
removal of the cutoff. Moreover we will assume
that $H$ and $H_\alpha$ satisfy the following properties, for a given $\delta>0$:

\noindent
(p1) \be
[H_\alpha,H_\beta]=0, \quad \forall \alpha, \beta\in I;
\label{412}
\en
(p2)\be
c_\alpha\equiv 2\delta \|H^{-1}\|\|H^{-1}H_\alpha\|<1, \quad \forall
\alpha\in I,
\label{413}
\en
and
\be
\lim_{\alpha\rightarrow 0}2 \|H^{-1}\|\|H^{-1}H_\alpha\|>0;
\label{414}
\en
(p3)
\be
\lim_{\alpha\rightarrow 0}\|[H-H_\alpha,Y]\|^{h,k}=0, \quad \forall Y \in
\Lc^+(\D);
\label{415}
\en
(p4)
\be
\|H^{-1}\|\leq \frac{1}{2(\delta+1)}.
\label{415bis}
\en
Before going on, we remark that conditions (p1) and (p3) together imply that
\be
[H,H_\alpha]=0, \quad \forall \alpha \in I.
\label{416}
\en

As discussed before, a typical problem in QM$_\infty$, consists in solving
first the
Heisenberg equations of motion \be
\frac{x_\alpha(t)}{dt}=i[H_\alpha,x_\alpha(t)], \quad \mbox{ with }
x_\alpha(0)=x
\label{417}
\en
for a  general observable $x$ in $\Lc^+(\D)[\tau]$, and, as a second
step, trying to remove the cutoff $\alpha$. This is equivalent to find the
$\tau$-limit of $x_\alpha(t)$ for
$\alpha$ going to zero.  In
this way we obtain the {\em dynamics} of the model and the time evolution of
$x$,
$x(t):=\tau-\lim_\alpha
x_\alpha(t)$.

As we know,
the solution of equation (\ref{417}) is, for finite $\alpha$,
\be
x_\alpha(t)=\exp{(iH_\alpha t)}x\exp{(-iH_\alpha t)}.
\label{418}
\en
Quite often, this expression is of little use since removing the cutoff in
(\ref{418}) is much
harder than working with the integral version of equation (\ref{417}):
\be
x_\alpha(t)=x+i\int_0^tds [H_\alpha,x_\alpha(s)].
\label{419}
\en
To analyze  the removal of the cutoff, we first define
\be F_\alpha(x_\eta(s))\equiv i[H_\alpha,x_\eta(s)], \quad \alpha, \eta\in I.
\label{420}
\en
Due to equations (\ref{412}) and (\ref{418}),we have: \be
F_\alpha(y_\eta(s))=\exp{(iH_\eta s)}F_\alpha(y) \exp{(-iH_\eta s)}\quad
\forall\alpha, \eta\in I
\mbox{ and } \forall y\in \Lc^+(\D). \label{421}
\en
It is convenient to introduce the set $\Lc^+_\gamma(\D)$ defined as
follows: we fix a value $\gamma$
in the set $I$;  $\Lc^+_\gamma(\D)$ is the set of
all the elements $y\in\Lc^+(\D)$ such that an element $y_0\in \Lc^+(\D)$
exists which satisfies
$y=\exp{(iH_\gamma t)}y_0\exp{(-iH_\gamma t)}$. Here both $y$ and $y_0$
could  depend on
time. Obviously, using equation (\ref{416}), it is easily
checked that $\Lc^+_\gamma(\D)$
is again $\tau-$complete. Moreover, it is also clear that
$\Lc^+_\gamma(\D)$ does not differ
significantly from $\Lc^+(\D)$ even from a purely algebraical point of
view. As a matter of
fact, its introduction is clearly only a technicality. We define on this
set the following map $U_\alpha$: \be
(U_\alpha y_\gamma)(t)\equiv x +\int_0^tds F_\alpha(y_\gamma(s)).
\label{422} \en Under the hypotheses
(p1)-(p4), and using the results of Section III, we will now prove that
$\{U_\alpha\}_{\alpha\in I}$ is
an uniform family of w$\tau$sc($\Lc^+_\gamma$), which is also $\tau$-strong
Cauchy. 

First, it is evident that each $U_\alpha$ is a w$\tau$sc($\Lc^+_\gamma$).

Secondly, taking $y_\gamma, z_\gamma \in \Lc^+_\gamma$, we have
$y_\gamma=e^{iH_\gamma t}y_t
e^{-iH_\gamma t}$ and $z_\gamma=e^{iH_\gamma t}z_te^{-iH_\gamma t}$, where
$y_t$ and $z_t$ belong
to $\Lc^+(\D)$ and could, in principle, depend on $t$.
Therefore, using equations (\ref{421}),
(\ref{416}) and the unitarity of the operators $\exp{(\pm iH_\gamma s)}$,
we get
\beano
\|(U_\alpha y_\gamma)(t)-(U_\alpha z_\gamma)(t)\|^{h,k} \leq
&&\!\!\!\!\!\int_0^t\|
\exp{(iH_\gamma s)} (F_\alpha(y_s)-F_\alpha(z_s)) \exp{(-iH_\gamma
s)}\|^{h,k}= \\
&&\!\!\!=\int_0^t\| (F_\alpha(y_s)-F_\alpha(z_s)) \|^{h,k}ds.
\enano
Since $\|(F_\alpha(y_s)-F_\alpha(z_s))
\|^{h,k}=\|h(H)[H_\alpha,y_s-z_s]H^k\|$, we have, inserting
twice $HH^{-1}$ in each term below:
\beano
 &&\!\!\!\!\!\!\|(F_\alpha(y_s)-F_\alpha(z_s)) \|^{h,k}\leq
\|h(H)H_\alpha(y_s-z_s)H^k\|+\|h(H)(y_s-z_s)H_\alpha H^k\|\leq \\
&&\!\!\!\!\!\!\leq 2\|H^{-1}\|\|H^{-1}H_\alpha\|\|y_s-z_s\|^{h_{+1},k+1}\leq
2\|H^{-1}\|\|H^{-1}H_\alpha\|\|y_\gamma(s)-z_\gamma(s)\|^{h_{+1},k+1}\leq \\
&&\!\!\!\!\!\!\leq
2\|H^{-1}\|\|H^{-1}H_\alpha\|\|y_\gamma-z_\gamma\|^{h_{+1},k+1}_\delta,
\enano
where we have used again the unitarity of the operators $\exp{(\pm
iH_\gamma s)}$ and the definition
of $\|\|^{h,k}_\delta$. Finally, definition (\ref{413}) gives
\be
 \|U_\alpha y_\gamma-U_\alpha z_\gamma\|^{h,k}_\delta \leq c_\alpha \|
y_\gamma-
z_\gamma\|^{h_{+1},k+1}_\delta.
\label{423}
\en
Of course, due to hypothesis (\ref{414}), we also get that
$$c_-=\lim_{\alpha \rightarrow 0}
c_\alpha = \delta \lim_{\alpha \rightarrow
0}2\|H^{-1}\|\|H^{-1}H_\alpha\|>0.$$ This is enough to
conclude that $\{U_\alpha\}_{\alpha\in I}$ is an uniform family of
w$\tau$sc($\Lc^+_\gamma$). To prove that it is also a $\tau$-strong Cauchy
net, we have to check that
$\|U_\alpha y_\gamma-U_\beta y_\gamma\|^{h,k}_\delta\rightarrow 0$ for all
$(h,k)\in \C_N$ and
for any $y_\gamma\in \Lc^+_\gamma$ when both $\alpha$ and $\beta$ go to zero.

Using the same procedure as above, we first obtain
$$
\|(U_\alpha y_\gamma)(t)-(U_\beta y_\gamma)(t)\|^{h,k}\leq \int_0^t
\|F_\alpha(y_s)-F_\beta(y_s)
\|^{h,k}ds. $$
This implies, after some easy estimates, that
$$
\|U_\alpha y_\gamma-U_\beta y_\gamma\|^{h,k}_\delta \leq \delta
\|[H_\alpha-H_\beta,y]\|^{h,k}_\delta, $$
and the rhs goes to zero because of the (\ref{415}).

Therefore, we conclude that Proposition 3 can be
applied. This means that the
dynamics for the model can be obtained as a $\tau$-limit of the regularized
dynamics, as obtained
from the equation (\ref{422}).

\vspace{3mm}

We end this Section, and the paper, with an explicit QM$_\infty$ model in
which conditions
(p1)-(p4) are satisfied. We refer to \cite{bag} for further details.

We take $\delta=1$. The starting point is the
pair of the annihilation
and creation operators $a$, $a^\dagger$, which satisfy the canonical
commutation relation
$[a,a^\dagger]=\Id$. Let $N=a^\dagger a$ be the number operator (which we
will identify with
its self-adjoint extension), with spectral decomposition
$N=\sum_{l=0}^\infty lE_l$. We take $N$ as the hamiltonian of the one mode free bosons. Of course, from the point of view of the dynamics, nothing 
change if
we add a constant to the hamiltonian. Therefore we define, for reasons
which will
be clear in the following, $H=4\Id+N=\sum_{l=0}^\infty(4+l)E_l$. As in
\cite{bag},
we introduce an occupation number cutoff $H\rightarrow H_L=4\Id+Q_LNQ_L$, where
$Q_L=\sum_{l=0}^L E_l$ is a projection operator. We can write
$H_L=\sum_{l=0}^\infty c_l^{(L)}E_l$, where $c_l^{(L)}$ is equal to $4+l$ for
$l=0,1,2,..,L$ and is equal to 4 for $l> L$. We also have
$H^{-1}=\sum_{l=0}^\infty(4+l)^{-1}E_l$ and
$H^{-1}H_L=H_LH^{-1}=\sum_{l=0}^\infty b_l^{(L)}E_l$, where
$b_l^{(L)}=\frac{c_l^{(L)}}{4+l}$.

Obviously we have:

$\bullet$ $[H_L,H_{L'}]=0, \quad \forall L,L'$;

$\bullet$ since $\|H^{-1}\|= \frac{1}{4} \left( \leq
\frac{1}{2(\delta+1)}=\frac{1}{4}\right)$ and
$\|H^{-1}H_L\|=1$, as it can be easily checked, then $$\lim_{L\rightarrow
\infty} 2
\|H^{-1}\|\|H^{-1}H_L\|=\frac{1}{2}>0;$$

$\bullet$ $\lim_{L\rightarrow \infty}\|[H-H_L,y]\|^{h,k}=0$. In fact we have
\beano
\|[H-H_L,y]\|^{h,k}&&\!\!\!\!\leq  \|(H-H_L)y\|^{h,k}+\|y(H-H_L)\|^{h,k}\leq \\
&&\!\!\!\!\leq\| \sqrt{h(H)}(H-H_L)\| \|
\sqrt{h(H)}yH^k\|+\|h(H)yH^{k+3}\|\|(H-H_L)H^{-3}\|.
\enano
We know that, if $h\in \C$ then also $\sqrt{h}\in \C$. Since the function
$h$ goes to zero
faster than any inverse power, using the spectral decompositions for $H$,
$H_L$ and $\sqrt{h(H)}$, it is easy to check that
$\lim_{L\rightarrow\infty}\|\sqrt{h(H)}(H-H_L)\|=0$.
Analogously it is not difficult to check that $\|(H-H_L)H^{-3}\|$ goes to 0 in the same
limit.

In this way we have checked that for the free bosons all the points of the
definition of a  uniform
family of w$\tau$sc $\tau$-strong Cauchy are satisfied, so that the
existence of the thermodynamical
limit of the model follows from the analysis
proposed in this paper.
\vspace{4mm}

\section{Concluding remarks}

In this paper we have discussed a possible extension of the notion of contraction map to a quasi *-algebraic framework, with
particular reference to the existence of fixed points and to the continuity of contractions
depending on a parameter. Both the mathematical and the physical
interest of the subject is, in
our opinion, quite evident. In particular, we believe that the possibility
of setting up a
new general approach for the problem of the existence of the dynamics for physical
problems in many-body
theory, quantum field theory or quantum statistical mechanics can be
considered as a nice result,
which deserves further studies. In particular, we believe that a deeper analysis of the set
$\B_L$ is certainly
worth. Also, a weakening of the hypotheses of Propositions 2 and 3 could be
relevant in order to
enlarge the class of models whose thermodynamical limit can be analyzed
following the procedure
proposed here. Finally, we plain to find additional conditions which ensure
uniqueness of the fixed
point and to consider the problem of the thermodynamical limit in the
Schr\"oedinger representation.

\vspace{4mm}

\noindent{\Large \bf Acknowledgments} \vspace{3mm}

It is a pleasure to thank C. Trapani for his stimulating suggestions.
This work has been supported by M.U.R.S.T.

\vspace{8mm}

\appendix
\renewcommand{\theequation}{\Alph{section}.\arabic{equation}}


 \section{\hspace{-14mm} Appendix:  On the non-triviality of $\B_L$}

The proof of the existence of non zero elements in the set $\B_L$ is better
carried out  working with
the topology $\tau_0$ we mentioned in Section I.

The building block for defining this new topology is a subset $\C_0$ of
 $\C$. We start introducing a fixed positive real number, $m$,  a finite subset of
$\N$, $J$, and a corresponding set of positive real numbers, $\{x_l, l\in
J\}$. Further, we
define \be \C_0\equiv \{f\in\C : f(x_l)\leq m, l\in J\}. \label{26}
\en

It is evident from this definition that $\C_0\subset\C$. It is also clear
that to any
function $f(x)\in\C$ can be associated, in a non-unique way, a function
$f_0(x)\in\C_0$ which
is proportional to $f(x)$. It is enough to take this proportionality
constant to be $$ \tilde
k=m \min_{l\in J'}(f(x_l)^{-1}), $$
where $J'$ is the largest subset of $J$ such that $ f(x_l)\neq O$ for all
$l\in J'$. (If
$ f(x_l)=O$ for all $l\in J$ then we can define $f_0(x)=0$.) We put
$f_0(x)=\tilde kf(x)$.

$\tau_0$ is the topology defined by the following seminorms

\be
X\in \Lc^+(\D) \rightarrow \|X\|^{f,k} \equiv
\max\left\{\|f(N)XN^k\|,\|N^kXf(N)\|\right\},
\label{27}
\en
where $k\geq 0$ and $f\in \C_0$. As for the topology $\tau$, we will
consider  only the
first term above, $\|f(N)XN^k\|$. It is evident that the two topologies
are indeed very close to each other. In fact, they are equivalent since the
above construction
implies that:

 $\bullet$ all the seminorms of the topology $\tau_0$ are also seminorms of
the topology $\tau$;

$\bullet$ all the seminorms of the topology $\tau$, $\|f(N)XN^k\|$, can be
written in terms of
a  seminorm of the topology $\tau_0$, $\tilde k \|f_0(N)XN^k\|$, where the
functions $f$ and
$f_0=\tilde k f$ belong respectively to $\C$ and $\C_0$.

For this reason  $\Lc^+(\D)$ turns out to be also $\tau_0$-complete. It  is
evident that the use of one or the other set of seminorms is completely
equivalent and that this choice is only a matter of convenience. For instance,
 the use of $\tau_0$ simplifies the
proof of the
non-triviality of the set  $\B_L$.

\vspace{3mm}

The first step of this proof consists in an analysis of the
spectrum of the operator $H$ involved in the definition of the seminorms.
We require to this
unbounded self-adjoint operator to have a spectrum with a discrete part and
with a
\underline{finite} number of eigenvalues $h_i$ with modulus not larger than
1. We call $\G$
the set of the corresponding indices: $|h_i|\leq 1, \forall i \in \G$. For
instance, if $N$ is
the usual number operator, whose spectrum is $\{0,1,2,3,4,...\}$, then the
operator $H=\frac{1}{5}N$
satisfies the above condition with $\G=\{1,2,3,4,5,6\}$.

We use now the set $\{h_i\}_{i\in\G}$ and a positive real $m$ to define the
set of functions $\C_0$
as  above and, by means of $\C_0$, the topology $\tau_0$.  
The role of the hamiltonian in the construction of the topology here is evident,
as in the original Lassner's paper, \cite{las}.

Calling $E_l$ the spectral projectors of the operator $H$, which from now
on will be assumed
to have discrete spectrum only to simplify the notation, we can write: $$
H=\sum_{l=0}^{\infty}h_lE_l=\sum_{l\in\G}h_lE_l +\sum_{l\not\in\G}h_lE_l,
$$
which implies that
$$
H^k=\sum_{l\in\G}h_l^kE_l +\sum_{l\not\in\G}h_l^kE_l, \quad h(H)=
\sum_{l\in\G}h(h_l)E_l +\sum_{l\not\in\G}h(h_l)E_l.
$$
Let now consider a set of complex number $\{c_i\}_{i\in\G}$ satisfying
condition
$\sum_{l\in\G}|c_l|\leq \sqrt{\frac{L}{m}}$, $L>0$. Starting with this set we
define the operator
$Y\equiv \sum_{l\in\G}c_l E_l$. Our aim is to show that the other operator
$X\equiv
Y^2=\sum_{l\in\G}c_l^2 E_l$ belongs to some $\B_{L'}$, at least under some
conditions on the w$\tau$sc $T$.

First of all we consider the following inequality
$\|X\|^{h,k}\leq \|h(H)Y\| \|YH^k\|.$
Secondly, we estimate separately the two contributions. We get
$$
\|YH^k\|= \|\sum_{l\in\G}c_l E_l\left[\sum_{l\in\G}h_l^kE_l
+\sum_{l\not\in\G}h_l^kE_l\right]\|\leq
\sum_{l\in\G}|c_l||h_l|^k\|E_l\|\leq \sum_{l\in\G}|c_l|,
$$
and, since $|h(h_l)|\leq m$ for all $l\in \G,$
$$
\|h(H)Y\|=\|\left[ \sum_{l\in\G}h(h_l)E_l +\sum_{l\not\in\G}h(h_l)E_l
\right] \sum_{l\in\G}c_l
E_l\|\leq \sum_{l\in\G}|c_l||h(h_l)|\|E_l\|\leq m \sum_{l\in\G}|c_l|. $$
Now, recalling the bound on the set $\{|c_i|\}_{i\in\G}$, we can conclude that
$$
\|X\|^{h,k}\leq m (\sum_{l\in\G}|c_l|)^2\leq L.
$$
Therefore, due to Lemma 1, point (a), we find that $X$ belongs to $\B_L$,
at least if $T0=0$.
We get a similar conclusion even in the weaker hypothesis on $T$ of Lemma
1, point (b).
Of course, it is not difficult to
generalize this strategy
in order to construct many other elements of $\B_L$.


\newpage

\begin{thebibliography}{99}

\bibitem{smart}D.R. Smart {\em Fixed Point Theorems},  Cambridge University
Press (1974)
\bibitem{rs}M.Reed and B.Simon, {\em Methods of Modern Mathematical
Physics}, I,
Academic Press, New York (1980)
\bibitem{kant}L.V. Kantorovich, G.P. Akilov {\em Functional Analysis in
Normed Spaces},  Pergamon Press (1964)
\bibitem{bm}F. Bagarello, G.
Morchio, {\em Dynamics of Mean-Field Spin Models from Basic Results in
Abstract Differential Equations}, J. Stat.  Phys. Vol.66, {\bf 3/4},
849-866 (1992)
\bibitem{hk} R. Haag, D. Kastler, {\em An algebraic approach to Quantum
Field Theory}, J. Math. Phys.,  {\bf 5}, 848-861, (1964)
\bibitem{las} G. Lassner, {\em Topological algebras and
        their applications in Quantum Statistics},  Wiss. Z. KMU-Leipzig,
     Math.-Naturwiss. R., {\bf 30}, 572-595, (1981), {\em Algebras  of
unbounded operators and quantum dynamics}, Physica, {\bf 124~A}, 471-480,
(1984)
\bibitem{ant} J.-P. Antoine, W. Karwowski, {\em Partial *-algebras of
closed linear operators in Hilbert space}, Publ.  RIMS Kyoto Univ., {\bf
21}, 205-236(1985)
\bibitem{btcq}F. Bagarello, C. Trapani, {\em States
and Representations of $CQ^*$-Algebras},  Ann. Inst. H. Poinc., {\bf 61},
103-133, (1994)
\bibitem{bag}F. Bagarello, {\em Applications of Topological *-Algebras of
Unbounded Operators}, J. Math.  Phys. Vol.66, {\bf 3/4},
849-866 (1992)
\bibitem{bag2}F. Bagarello, {\em Applications of Topological *-Algebras of
Unbounded Operators to Modified Quons}, submitted to Publ. RIMS
\bibitem{bt}F. Bagarello, C. Trapani, {\em "Almost" Mean
Field Ising Model: An Algebraic Approach}, J. Stat. Phys. Vol.65, {\bf
3/4}, 469-482 (1991)
\bibitem{bt2}F. Bagarello, C. Trapani, {\em The Heisenberg Dynamics of
Spin Systems: A Quasi$^*$-Algebras Approach}, J. Math. Phys., {\bf
37}, 4219-4234 (1996)
(1955) 




\end{thebibliography}
\end{document}